\documentclass[12pt]{article}
\usepackage{epsf}

\newcommand {\be}{\begin{equation}}
\newcommand {\ee}{\end {equation}}

\begin{document}

\hspace{10cm}{\bf NT@UW-04-06}\\

\begin{center}

{\bf Corrections to the Nuclear Axial Vector Coupling in a Nuclear Medium}
\end{center}
\begin{center}
{G. W. Carter$^a$ and E. M. Henley$^{a,b}$  }\\
{\small \em $^{a}$Department of Physics, Box 351560, University of Washington,\\
Seattle, WA 98195-1560, USA }\\
{\small \em $^{b}$Institute for Nuclear Theory, Box 351550,
University of Washington, \\
Seattle, WA 98195-1550, USA }\\
\end{center}
\vspace{0.25 in}
\begin{abstract}
There is experimental evidence for an increase in the temporal component 
of the weak axial vector current in nuclei.
Most of this increase is accounted for by meson exchange currents, and we consider
further corrections from pions and the sigma mean-field.
These corrections, however, are found to reduce the alreadly insufficiently large coupling.
\end{abstract}

\vspace {0.25 in}
\section{Introduction}
Unlike the vector current, $j_\mu$, the weak axial current $j_\mu^5$ is not 
conserved. In a nuclear medium or nucleus, relativistic invariance is lost
and space and time components become disconnected. Experimentally, $ g_A 
\bar{N}
\mathbf{\gamma} \gamma^5 N \sim g_A \bar{N}
{\mathbf{\sigma}} N$ appears to decrease from its free space value of 1.26 
by about 10\% \cite{1}. A variety of reasons have been given for this decrease,
including nuclear structure effects, the role of the 
$\Delta (1232)$ \cite{Park},
chiral symmetry restoration \cite{CDE}, and 
short-range correlations \cite{LTT}. 
Carter and Prakash \cite{CP}, for example, used a chiral Lagrangian with a mean field
$\sigma$ to find $g_A^* \sim 1.1$ in neutron star matter at normal nuclear density.

The time component of the axial vector current can be obtained from 
first-forbidden beta decays \cite{W}, e.g.in $0^- \rightarrow 0^+$
transitions, or in allowed
Gamow-Teller decays (e.g. $1^+ \rightarrow 0^+$) with polarized nuclei
\cite{MM}. The angular asymmetry from polarized mirror nuclei is only sensitive to
$g_A^{0*} \bar{N} \gamma^0 \gamma^5 N$. 

\begin{figure}[tb]
\centerline{
\setlength\epsfxsize{18mm}
\epsfbox{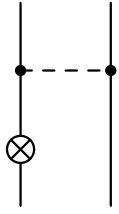}
}
\caption{
Pion exchange (dashed line) between nucleons (solid lines).
The large crossed circle is the bare axial current coupling.
}\label{exch_fig}
\end{figure}

Already in the 1970's Kubodera, Delorme, and Rho \cite{KDR} showed that 
nuclear exchange currents are responsible for, at least, the major part of the 
increase of $g_A^{0*}$. Why are exchange currents so important in the time 
component and not for the space part? The reason is that the pion adds an 
extra $\gamma^5$ (see Fig.~\ref{exch_fig})  which makes the space component
effectively $\bar{N} {\mathbf{\gamma}} \gamma^5 \gamma^5 N = \bar{N}
{\mathbf{\gamma}} N \sim \bar{N} {\cal O}\left(\frac{{\bf{p}}}{M}\right) N$, 
whereas the 
time component becomes $\bar{N} \gamma^0 \gamma^5 \gamma^5 N
= \bar{N} \gamma^0 N \sim \bar {N} {\cal O}(1) N$. Thus pion exchange 
currents are more important for the temporal component. By contrast, this
component of the axial current do not couple to the $\Delta(1232)$ whereas the
space component does.

For the time component of the axial vector current, pion exchange currents
have been examined by several authors \cite{W,KDR,KR,T}; some
have also included shorter range exchanges such as two pions, $\rho$, and $A_1$. 
Some aspects of the exclusion principle have been included in pion exchange calculations, 
e.g., via use of the shell model \cite {W,T}. The authors
find, approximately,
\be
g_A^{0*} = g_A (1 + \delta)\,,
\ee
with $\delta $ approximately given by
\begin{eqnarray}
A \sim 16 & \delta_{exp} \approx 0.64 & \delta_{th} \approx 0.5-0.6 \,,
\nonumber\\
A \sim 96 & \delta_{exp} \approx 0.75 & \delta_{th} \approx 0.6 \,, \nonumber\\
A \sim 200 & \delta_{exp} \approx 1  & \delta_{th}  \approx 0.8 \,. \nonumber 
\end{eqnarray}

In addition to the normal exchange current contributionbs to $g_A^{0*}$, there is,
to the same order, the one nucleon pion loop contribution. It differs from the 
free nucleon counterpart by the effect of the Pauli exclusion principle.
It is this effect we have examined in order to see whether it might 
contribute the missing $\sim 10 \%$ of $g_A^{0*}$. 

\section {Theory}

We start with the chiral Lagrangian of Carter et. al. \cite{CER,CP}, with 
the relevant parts
\begin {eqnarray}
\emph{L}_1 &=& - \bar{N}g \left[ (\sigma + i \vec{\tau} \cdot \vec{\pi}
\gamma_5) + 
\frac{1}{2}g' \gamma^\mu \vec{\tau} \cdot \vec{a}_\mu \gamma_5\right] 
N \,,\\
\emph{L}_2 &=& \frac{1}{2} \frac{D}{\sigma_0^2} \bar{N} \left[ \gamma^\mu
\vec{\tau} 
\cdot[\vec{\pi} \times \Delta_\mu \vec{\pi} + \gamma_5 (\sigma \Delta_\mu
\vec{\
\pi} - \vec{\pi} \Delta_\mu \sigma)]\right] N \,,\nonumber \\
\Delta_\mu \sigma &=& \partial _\mu \sigma + g' \vec{a_\mu} \cdot \vec{\pi}
\,, \\
\Delta_\mu \vec{\pi} &\approx& \partial _\mu \vec {\pi} - g' \sigma
\vec{a}_\mu \,,
\end {eqnarray}
where $g'$ is the coupling to the chiral vector mesons $\rho$ and $A_1$, and
arrows indicate isospin vectors.

\begin{figure}[tb]
\centerline{
\setlength\epsfxsize{80mm}
\epsfbox{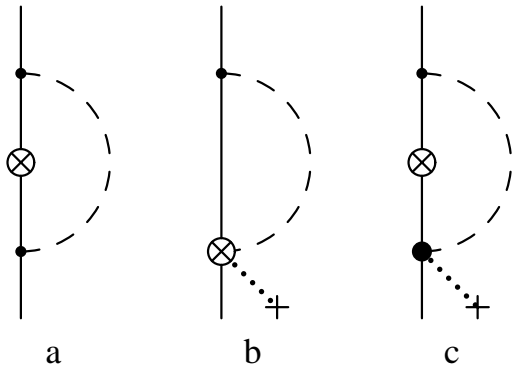}
}
\caption{
Leading contributions to the axial coupling in medium.  Solid, dashed, and 
dotted lines respectively denote nucleons, pions, and sigma mesons (which
terminate in the chiral mean field).  The large crossed circles are bare
axial current couplings.
}\label{abc_figs}
\end{figure} 

As in the previous works \cite {CER,CP}, we treat the $\sigma$ field in mean field theory. 
The diagrams shown in Fig.~\ref{abc_figs} contribute. Consider Fig~\ref{abc_figs}a. 
To start with,  the nucleon has a
momentum  $ \leq k_F$. Since we need the propagators in the presence of the 
nucleus, they have to be above the Fermi sea (e.g. ${\bf|{p}|} > k_F$), 
and we have
\be
\frac{1}{
\not\!p - M + i \epsilon} \rightarrow \frac{(\not\!p +
M)\Theta(|{\bf{p}}|
- k_F)}{2 E_p (p_0 - E_p + i \epsilon)} \,.
\ee
Here $E_p =  \sqrt{{\bf{p}}^2 + M^2}$.

The effect of the in-medium sigma field is to reduce the nucleon mass, 
$M \rightarrow M^* \sim 0.8 M$.
We need to compute integrals of the form
$\int_{k_F}^\infty d^3p F(p,M^*)$, but only want the change of
$g_A$ from the free nucleon value. We thus need
\be
\int_0^\infty  d^{3}p [F(p,M^*) - F(p,M)] - \int_0^{k_F} d^3p F(p,M^*) 
\equiv \left({\cal I}^*_\infty - {\cal I}_\infty\right) - {\cal I}^*_{k_F}\,,
\ee
where we have defined
${\cal I}^*_k = \int_0^k d^3p F(p,M^*)$
for integrands $F(p,M^*)$ corresponding to Figs.~\ref{abc_figs}.
The last integral is always finite, but not so the first ones. 
We thus assume that the pion coupling to the nucleon has a dipole form factor
\be
f(k,\Lambda) = \left(\frac{\Lambda^2}{ \Lambda^2 + k^2}\right)^2\, ,
\ee
with the cutoff $\Lambda$ = 0.9 or 1.1 GeV.

The constant D was fixed in Ref.~\cite{CER} to fit $g_A= 1.26$
for the free nucleon,
\be
g_A \approx \left(1 + D \frac{\bar{\sigma}^2}{\sigma_0^2}\right)\,,
\ee
where $\sigma_0$ is the vacuum expectation value of the $\sigma$ field, 
$\sigma_0 = 102$ MeV, and $\bar{\sigma}$ is the mean field result at
finite density. There are 
corrections due to heavier meson fields, which we have omitted here. The
reduction of the space part of the axial vector coupling in a medium occurs due to 
to the decrease of $\bar{\sigma}^2$. This reduction 
occurs for the time component as well. 

\section{Results}

In our work there is a need to readjust the value of the constant D due to 
the inclusion of higher order diagrams. This reduces D by about 10\%. 
Results are shown in the Table 1.  We
find that the contribution of Fig.~\ref{abc_figs}b 
is large and negative  (i.e., $\delta < 0$).
Furthermore the difference between the contribution of the $M^*$ and $M$ terms 
is negative for Fig.~\ref{abc_figs}a 
and almost cancels the contribution due to the 
integral up to $k_F$. 
The contribution of Fig.~\ref{abc_figs}c is positive but small There also
two loop diagrams and other higher order terms; they give a negligible contribution. 

\begin{table}[t]
\begin{center}
\begin {tabular}{|l l|ll|} \hline
  &              & $\Lambda$= 1.1 GeV    & $\Lambda$ = 0.9 GeV \\ \hline
Fig.~\ref{abc_figs}a & ${\cal I}^*_{k_F}$ &+.21                  & +.19 \\
  &     ${\cal I}^*_\infty - {\cal I}_\infty$ & -.26  & -.21 \\ \hline
Fig.~\ref{abc_figs}b & ${\cal I}^*_{k_F}$ & -.18                  & -.17 \\
  &     ${\cal I}^*_\infty - {\cal I}_\infty$   & -.03    & -.02\\ \hline
Fig.~\ref{abc_figs}c & ${\cal I}^*_{k_F}$ & +.01                  & +.004 \\
  &     ${\cal I}^*_\infty - {\cal I}_\infty$   & +.08    & +.054 \\ \hline
\end{tabular}
\caption{
Corrections to the axial coupling at nuclear matter density
corresponding to the diagrams of Fig.~2.
}
\end{center}
\end{table}

In sum, $g_A^* = g_A (1 + \delta)$, with $\delta = -.13 (-.11)$ for 
$\Lambda = 1.1 (0.9)$ GeV.
The omitted effect is indeed of the order of magnitude 
required (10-15\%), but with the wrong sign. These corrections only serve
to increase the discrepancy between theory and experiment.

The authors thank Mary Alberg for helpful comments.  This work was supported
by the U. S. Department of Energy grant DE-FG02-97ER4014.

\end{document}